\documentclass[aps,prl,superscriptaddress,showpacs,onecolumn,floatfix,amsmath,amssymb]{revtex4-2}

\RequirePackage{fix-cm}
\usepackage{graphicx}
\def\gtorder{\mathrel{\raise.3ex\hbox{$>$}\mkern-14mu
 \lower0.6ex\hbox{$\sim$}}}
\def\ltorder{\mathrel{\raise.3ex\hbox{$<$}\mkern-14mu
 \lower0.6ex\hbox{$\sim$}}}

\begin{document}

\title{PVEMC: Isolating the flavor-dependent EMC effect using parity-violating inelastic scattering in SoLID}

%arXiv \titlerunning{PVEMC: Flavor dependence of the EMC effect}    

\author{Rakitha Beminiwattha}  
\affiliation{Louisiana Tech University, Ruston, Louisiana, 71272, USA}
\author{John Arrington}  
\affiliation{Lawrence Berkeley National Laboratory, Berkeley, California 94720, USA}
\author{David J. Gaskell}
\affiliation{Thomas Jefferson National Accelerator Facility, Newport News, Virginia, 23606}

%arXiv \institute{R. Beminiwattha \at
%arXiv               Louisiana Tech University, Ruston, Louisiana, 71272, USA \email{rakithab@latech.edu}
%arXiv        \and
%arXiv             J. Arrington \at
%arXiv               Lawrence Berkeley National Laboratory, Berkeley, California 94720, USA 
%arXiv            \and
%arXiv            D. Gaskell \at
%arXiv               Jefferson Lab, Newport News, Virginia, 23606  \\
%arXiv }

%arXiv \date{Received: date / Accepted: date}

%arXiv \maketitle

\begin{abstract} 
In order to better understand the EMC effect, we propose a clean and precise measurement of the flavor dependence of the EMC effect using parity-violating deep inelastic scattering on a $^{48}$Ca target. This measurement will provide an extremely sensitive test for flavor dependence in the modification of nuclear parton distribution functions (PDFs) for neutron-rich nuclei. A measurement of the flavor dependence will provide new and vital information and help to explain nucleon modification at the quark level. In addition to helping understand the origin of the EMC effect, a flavor-dependent nuclear pdf modification could significantly impact a range of processes, including neutrino-nucleus scattering, nuclear Drell-Yan processes, and e-A observables at the Electron-Ion Collider.

The parity-violating asymmetry $A_{PV}$ from $^{48}$Ca using an 11 GeV beam at $80~\mu A$ will be measured using the SoLID detector, proposed for a series of measurements in Hall A at Jefferson Lab. In 68 days of data taking, we will reach $0.7-1.3\%$ statistical precision for $0.2<x<0.7$ with $0.6-0.7\% $ systematic uncertainties.  The goal is to make the first direct measurement of the flavor dependence of the EMC effect.  The precision of the measurement will allow for quantification of the flavor-dependent effects, greatly improving our ability to differentiate between models of the EMC effect and constraining the u- and d-quark contributions in neutron rich nuclei. 

\end{abstract}
\maketitle

\section{Introduction} \label{sec:intro}

An open and important question for hadronic physics today is how protons and neutrons are modified when they are bound in a nucleus. The observation of the ``EMC effect'', the depletion of the nuclear quark distributions for $0.3<x<0.8$ relative to the expectation from nucleon parton distribution functions (PDFs) plus Fermi motion, provides clear evidence that the nuclear pdfs are not simply the sum of the pdfs of unmodified proton and neutrons~\cite{Geesaman:1995yd,Norton:2003cb,Malace:2014uea}. But despite this direct measurement of such modification, the underlying physics mechanism(s) for it is not well understood. While the existence of nuclear modification of the pdfs is well established, important questions remain about the nature of the modification: a detailed description of its A dependence is not yet complete, and we have almost no experimental information on its spin- and flavor-dependence. An improved understanding of these questions will be enormously important in guiding a theoretical understanding, but it is also a pressing experimental issue with broad implications.

\subsection{PVEMC overview}
The scattering cross section for electroweak neutral current is dependent on both the amplitudes for the exchanged virtual photon and neutral $Z$ boson, $\sigma \propto \left| A_\gamma + A_Z \right|^2$ where $A_\gamma, A_Z$ are photon and $Z$ boson scattering amplitudes, respectively. The dominant term for the cross section is $|A_\gamma|^2$ and the interference term $|A_\gamma^*A_Z|$ is a non-zero parity violating term for $Q^2 \ll M_Z^2$. While absolute cross-section dominated by $A_\gamma$ for $Q^2 \ll M_Z^2$, one can form a parity-violating quantity which is measured by the differences between right and left-handed polarized lepton absolute cross sections for $Q^2 \ll M_Z^2$,
\begin{equation}
    A_\mathrm{PV} = \frac{ \sigma_R - \sigma_L }{\sigma_R + \sigma_L} \sim \frac{|A_\gamma^*A_Z|}{|A_\gamma|^2}. 
    \label{eq:int:apv}
\end{equation}
At sufficiently large momentum and energy transfer from an electromagnetic probe to a hadronic target, a transition takes place where the underlying QCD degrees of freedom are exposed. In this deep inelastic scattering (DIS) region, the target appears as an incoherent sum of weakly interacting partons which we identify as quarks. At this scale this asymmetry provides a particularly sensitive method to obtain flavor-dependent effects in nuclear modification as it is a ratio of the weak-to-electromagnetic interactions, which gives access to ratios of quark distributions.
In the quark-parton model, the right-left polarized lepton scattering asymmetry given in Eq.~\ref{eq:int:apv}  can be expressed in terms of the parton distribution functions by
\begin{equation}
    A_\mathrm{PV} = \frac{G_F Q^2}{4 \sqrt{2} \pi \alpha} \left[ Y_1 a_1(x) + Y_3(y) a_3(x) \right],
    \label{eq:phy:apv}
\end{equation}
where $G_F$ is the Fermi constant and $\alpha$ is fine structure constant. 
\begin{equation}
    Y_1 \approx 1 ~~;~~ Y_3(y) \approx \frac{1 - (1-y)^2}{1 + (1-y)^2}
\end{equation}
where  $e_i$ is quark's electric charge, $q_i$ and $\bar q_i$ are quark/anti-quark PDFs with 
\begin{eqnarray}
    a_1(x) = g_A^e \frac{F_1^{\gamma Z}}{F_1^\gamma} = 2 \frac{ \sum_i C_{1i} e_i (q_i + \bar{q}_i) }{ \sum_i e_i^2 (q_i + \bar{q}_i) } ; 
    a_3(x) = g_V^e \frac{F_3^{\gamma Z}}{2 F_1^\gamma} = 2 \frac{ \sum_i C_{2i} e_i (q_i - \bar{q}_i) }{ \sum_i e_i^2 (q_i + \bar{q}_i) }. 
\end{eqnarray}
with $ \nu=E-E' $, $y=\nu/E$, $g_A^e$ and $g_V^e$ are electron axial and vector couplings to the $Z^0$ boson, respectively, and $C_{1i}$ and $C_{2i}$ are the effective quark ($i=u$ and $d$, respectively) couplings constants. In practice, the $a_1$ term dominates the asymmetry as the $C_{2i}  = \mp \frac{1}{2} \pm 2 \sin^2\theta_W$ couplings are suppressed by an order of magnitude relative to $C_{1i}$.
$C_{1i}$ is given in terms of $\sin^2\theta_W$ as below
\begin{equation}
\label{eqn:c1ud}
C_{1u} = -\frac{1}{2} + \frac{4}{3} \sin^2\theta_W, ~~~~ 
C_{1d} = \frac{1}{2} - \frac{2}{3} \sin^2\theta_W. 
\end{equation} 
The power of this method is elucidated when you start from the asymmetry for the case of an isoscalar nucleus at large enough $x$ that the sea quarks do not contribute significantly,
\begin{equation}
    a_1 \simeq \frac{9}{5} - 4\sin^2\theta_W~~,
\end{equation}
and then examine the impact of a small difference between the up- and down-quark distributions (to first order in the light-quark asymmetry, $u_A - d_A$):
\begin{equation}
    \label{eqn:a1-flav-dep}
    a_1 \simeq \frac{9}{5} - 4\sin^2\theta_W - \frac{12}{25} \frac{u_A^+ - d_A^+}{u_A^+ + d_A^+}  ~~,
\end{equation}
with the convention that $q^\pm = q(x) \pm \bar{q}(x)$ and using the definitions of $C_{1i}$ in eq.~\ref{eqn:c1ud}. Parity-violating deep inelastic asymmetry measurements are therefore directly sensitive to differences in the quark flavors.  In turn, for isoscalar targets in these kinematics, $a_1$ roughly becomes a constant and the measurement becomes a test for charge symmetry violation~\cite{Londergan:2009kj}.

\subsection{EMC effect overview}

\begin{figure}[htb]
    \begin{center}
        \includegraphics[width=0.75\textwidth,trim={0mm 0mm 18mm 18mm}, clip]{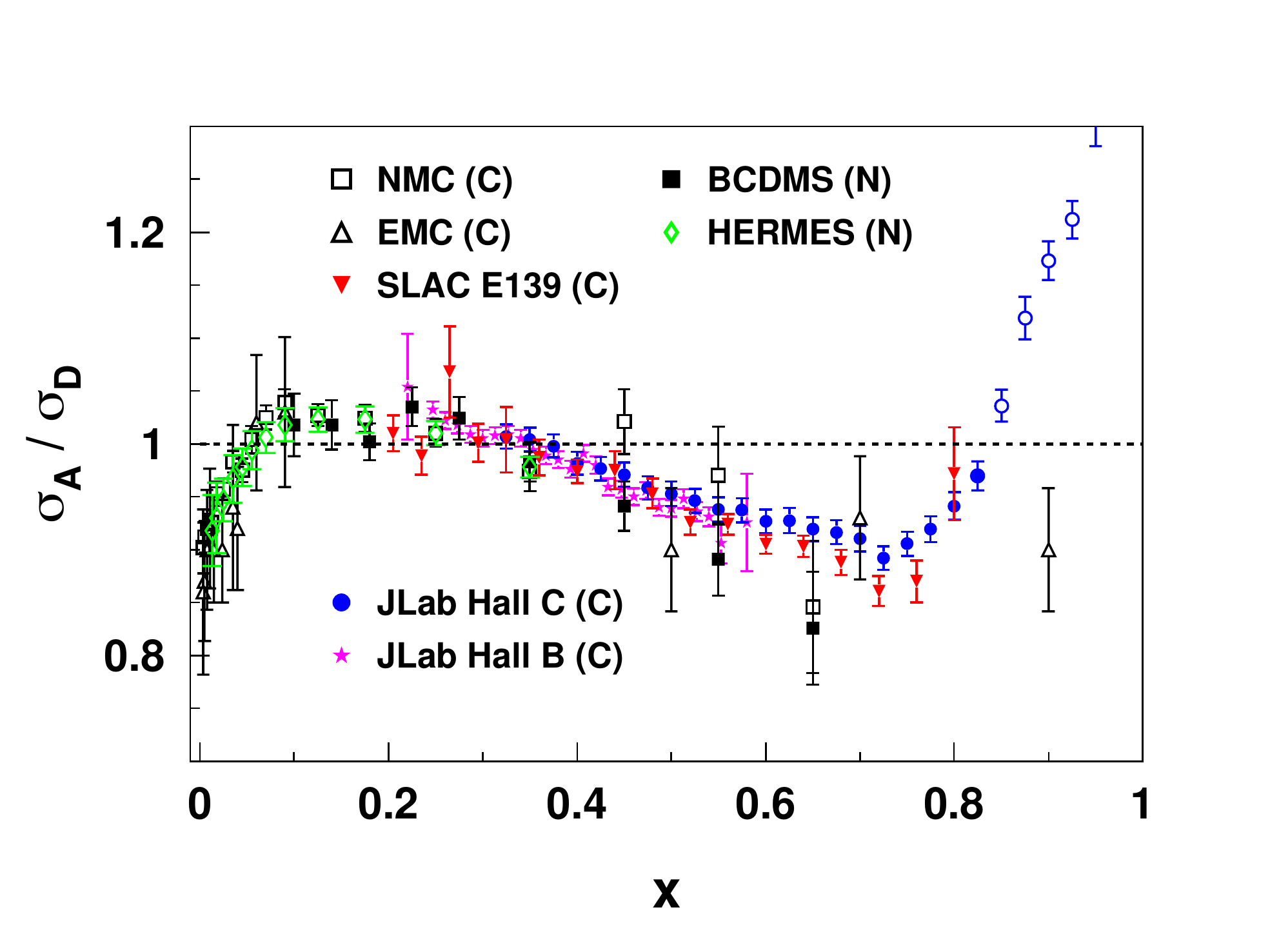}       
        \caption{Measurement of the EMC effect in carbon and nitrogen from EMC~\cite{EuropeanMuon:1989mef}, BCDMS~\cite{BCDMS:1985dor}, NMC~\cite{NewMuon:1995tgs}, SLAC~\cite{Gomez:1993ri}, Hermes~\cite{HERMES:1999bwb}, and JLab Hall C~\cite{Arrington:2021vuu} and Hall B~\cite{CLAS:2019vsb}.}
        \label{fig:unpolarized_emc}
    \end{center}
\end{figure}

The European Muon Collaboration made the first (unexpected) observation of the modification of inelastic structure functions in a nucleus in 1983~\cite{EuropeanMuon:1983wih}. This was the first unambiguous signature of nuclear effects in deep inelastic scattering, beyond those expected from simple Fermi motion, and resulted in a significant re-examination of the pre-conceived notion that the energy scales involved in nuclear binding ($\sim$MeV) should not impact the dynamics of GeV-scale probes used in measurements of deep inelastic scattering (DIS). Since the initial observation, there have been several subsequent measurements elucidating the properties of what became known as the EMC effect - the deviation of the A/D cross section ratio from unity.  These measurements provided data with electron and muon beams, over a wide range of beam energies (from a few to hundreds of GeV), and for a variety of targets (from $A=3$ to $A=208$).  See Ref.~\cite{Malace:2014uea} for a review of available data.  This significant body of data has resulted in a few notable observations:
\begin{itemize}
 \item{The shape of the EMC effect, shown in Fig.~\ref{fig:unpolarized_emc}, shows a suppression at low $x$ (shadowing), a small enhancement in the region $0.1<x<0.3$ (anti-shadowing), a suppression between $0.3<x<0.8$ (the EMC region), and an increase at large $x$ due to the effects of Fermi motion~\cite{Geesaman:1995yd}.}
 \item{The EMC ratio appears to be nearly independent of $Q^2$.}
 \item{The size of the EMC effect increases slowly with $A$ for A$>$12, but has more complicated structure for lighter nuclei~\cite{Gomez:1993ri,Seely:2009gt,Arrington:2021vuu,HallC:2022rzv}.}
\end{itemize}
The last observation has resulted in new questions on nuclear dependence of the EMC effect, i.e., whether the effect depends on $A$, nuclear density, or some other independent variable~\cite{Arrington:2012ax}.

There have been a variety of ideas proposed to explain the origin of the EMC effect, including contributions due to nuclear binding, the nuclear "pion excess" from enhancement of the nucleon virtual pion cloud due to pion exchange between nucleons, dynamical rescaling of structure functions in the nucleus, and multi-quark clusters (see Refs.~\cite{Malace:2014uea,Arrington:2021vuu} more detailed discussions). A more recent approach relies on effects due to quark interactions with meson fields generated by other nucleons~\cite{Cloet:2009qs}. Notably, this model predicts a significant EMC effect in spin-dependent structure functions~\cite{Cloet:2006bq} as well as a dependence on the flavor of the quarks. Several of these ideas will be examined at Jefferson Lab over the next 5-10 years~\cite{Arrington:2021alx}. This article focuses on the question of a possible flavor dependence to the EMC effect. Until recently, nearly all models of the EMC effect and most extractions from nuclear pdf analyses implicitly assumed that both up and down quarks experience the same modification in a nucleus and have no mechanism to allow a difference. One exception would be models that assume that the EMC effect is driven primarily by nucleons found in short-range correlated pairs which are dominated by np pairs~\cite{Arrington:2022sov}.

\subsection{Flavor dependence of the EMC effect}

Over the past decade or so, the question of the flavor dependence of the EMC effect has become a topic of renewed interest~\cite{Arrington:2015wja,Cloet:2019mql}. Calculations~\cite{Cloet:2012td}, simple scaling models~\cite{Arrington:2015wja}, and experimental observations~\cite{Seely:2009gt,Arrington:2012ax,Fomin:2011ng} suggest that the EMC effect should be flavor dependent, especially for non-isoscalar nuclei. These effects are generally much larger than the very small effects that can arise from conventional, isospin-independent physics such as Fermi smearing. Even for the case of identical proton and neutron momentum distributions, Fermi smearing has a slightly different impact on the proton and neutron due to the difference in the pdfs for the free proton and neutron. Flavor dependent effects will impact the nuclear PDFs of heavy neutron-rich nuclei and the nuclear PDFs need to be well understood for precision lepton scattering, as measurements involving p-A and A-A collisions that are sensitive to both the u- and d-quark distributions. 

The observation of a non-trivial correlation~\cite{Arrington:2012ax} between the presence of short-range correlations (SRCs)~\cite{Fomin:2011ng} and the EMC effect~\cite{Seely:2009gt}, combined with the near-universal dominance of np pairs in SRCs~\cite{Arrington:2022sov,Li:2022fhh}, suggest the possibility of flavor dependence of the EMC effect~\cite{Arrington:2012ax,CLAS:2019vsb,Arrington:2021vuu}. This depends on the idea that the correlation is causal, with the high-momentum component of SRCs yielding modification of the nuclear PDFs, as opposed to both the EMC effect and SRCs being driven by a common origin, e.g. simple nuclear density, with the isospin dependence of SRCs generated by the isospin structure of the NN potential~\cite{Schiavilla:2006xx,Alvioli:2007zz}.  Examinations of the quality of the EMC-SRC correlation, comparing causal and non-causal interpretations~\cite{Arrington:2012ax} do not show a significant preference for either of these approaches.  Similarly, examinations of the quality of universal scaling of the EMC effect with the presence of SRCs~\cite{Arrington:2019wky} are also unable to provide meaningful sensitivity to the question of whether or not the EMC effect needs to be flavor dependent.

While there is no direct experimental evidence for a flavor dependence of the EMC effect, we can use models and simple scaling assumptions to predict the potential size of the flavor dependence.  Ref.~\cite{Cloet:2012td} calculates the EMC effect for up- and down-quark contributions separately, yielding a prediction of both the $x$ and nuclear dependence of the flavor dependence of the EMC effect, at least for medium-to-heavy nuclei. Because the flavor dependence is generated by the scalar and vector QCD potential, the flavor dependence arises only in nuclei that have different proton and neutron numbers.  

It is also possible to take simple assumptions about the scaling behavior of the EMC effect in nuclei and use precise \textit{ab initio} calculations of nuclear structure~\cite{Wiringa:2013ala} to determine the scaling of the EMC effect in light nuclei. Such assumptions can include the EMC effect scaling with average nuclear density, `local' density (related to the probability for nucleons to be close enough together to have significant overlap), or average virtuality or other measures of the contribution from high-momentum nucleons. These can be calculated for a range of light nuclei, and can also be evaluated separately for proton and neutron in these nuclei, yielding an estimate of the fractional difference between proton and neutron EMC effect~\cite{Arrington:2015wja}.  

We note that for all of the assumptions described above: a causal relationship between the EMC effect and SRCs, the direct calculation of the flavor dependence, or any of the simple scaling models, the effect is predicted to be larger for protons in neutron-rich nuclei, although the size of the flavor dependence in these different approaches can vary significantly.  In later sections, we will use these models to make predictions for the parity-violating asymmetry and evaluate the sensitivity of this measurement to the flavor dependence of the EMC effect.

\section{\label{sec:flavmeas}The Proposed PVEMC Measurement of the Flavor Dependence of the EMC effect }
Parity-violating deep inelastic asymmetry measurements are directly sensitive to differences in the quark flavors as illustrated by Eq.~\ref{eqn:a1-flav-dep}. To examine this hypothesis that there is a significant flavor dependence in non-isoscalar nuclei, we proposed to utilize large proton-neutron asymmetry nuclei. For this experiment, we choose a target of $^{48}$Ca. $^{48}$Ca has a larger fractional neutron excess than other nuclei of similar mass, and as such is expected to have a larger flavor dependence, as seen in the Clo\"{e}t-Bentz-Thomas (CBT) model~\cite{Cloet:2012td}, described below, for which the flavor-dependent effect is half the size in $^{56}$Fe compared to $^{48}$Ca. For heavier nuclei like $^{208}$Pb (or targets like depleted uranium and gold with similar $N/Z$), the flavor dependence scales roughly as $N/A$. While this yields a $\sim$25\% larger effect than for $^{48}$Ca, these high-Z nuclei have much larger radiative and Coulomb corrections, and generate much larger radiation backgrounds at the same thickness as the $^{48}$Ca target. The loss of rate associated with maintaining acceptable radiation levels is large enough that the overall figure of merit is worse, even with the larger expected signal for flavor dependence.

We will discuss a range of models that can make quantitative predictions of the flavor dependence of the EMC effect, which can provide guidance as to the expected signal size and thus inform the precision required for making a significant measurement. However, the goal of the experiment is not to test any specific model, but to provide a clean and sensitive extraction of flavor-dependent nuclear effects in a sector where other potential measurements have limited sensitivity or significant model dependence.  

\subsubsection{Clo\"{e}t-Bentz-Thomas (CBT) model} \label{sec:phys:cloet}
It was proposed by Clo\"{e}t, Bentz, and Thomas~\cite{Cloet:2009qs} that one possible resolution to the NuTeV~\cite{NuTeV:2001whx} anomaly was through the existence of an isovector EMC effect.  These calculations, referred to here as the CBT model, evaluate the impact of the scalar and vector mean field of the nucleus on the nucleon pdfs within the Nambu-Jona-Lasinio Model~\cite{Nambu:1961tp}. 

\begin{figure}[htb]
    \begin{center}
    \includegraphics[width=0.325\textwidth,trim={0 0 18mm 18mm}, clip]{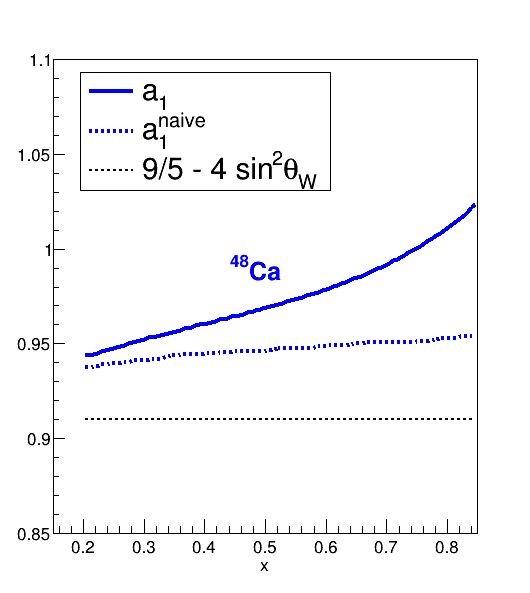}
    \includegraphics[width=0.325\textwidth,trim={0 0 18mm 18mm}, clip]{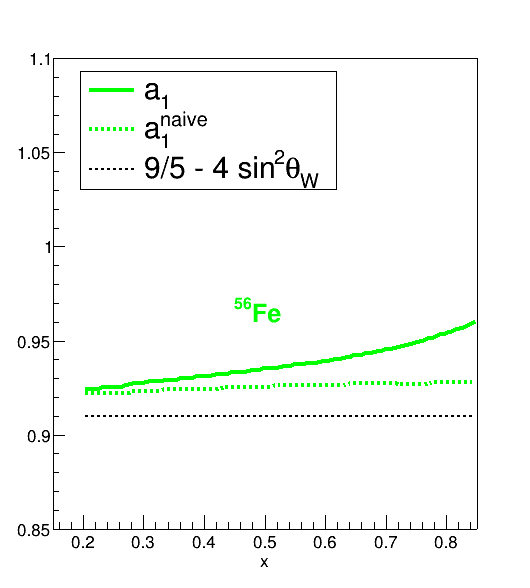}
    \includegraphics[width=0.325\textwidth,trim={0 0 18mm 18mm}, clip]{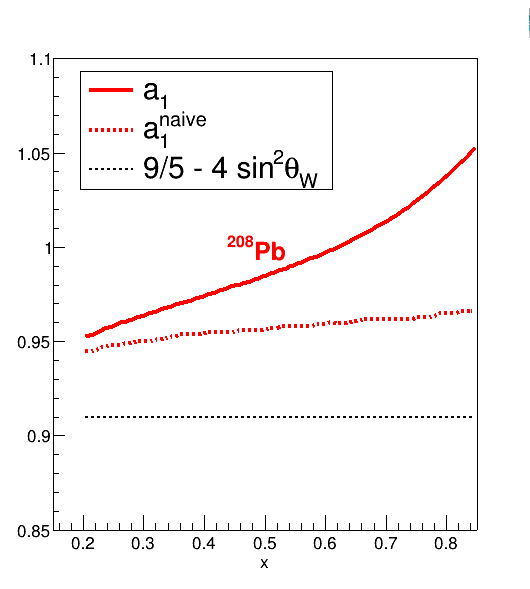}
        \caption{CBT model $a_1$ predictions for $^{48}$Ca, $^{56}$Fe, and $^{208}$Pb from~\cite{Cloet:2012td} (solid colored lines) compared to the prediction assuming that the EMC effect is flavor independent (dotted colored line). The black dotted line indicates the expectation for an isoscalar nucleus.}
        \label{fig:phy:cloetinterp_reprod}
    \end{center}
\end{figure}

The CBT model model has been successful in reproducing the quark distributions for the EMC effect and the measured structure functions. Predictions are made within this model for the PVDIS $a_1$ term for $^{56}$Fe, $^{48}$Ca, and $^{208}$Pb and are shown in Fig.~\ref{fig:phy:cloetinterp_reprod}. The impact of the flavor-dependence grows with $N/Z$, yielding the largest effects for $^{208}$Pb and $^{48}$Ca, and a much smaller effect for $^{56}$Fe. The calculation shows a clear enhancement in $a_1$ over the prediction for a flavor-independent EMC effect, labeled $a_1^{naive}$, that grows with increasing $x$. There is essentially no difference at $x=0.2$, and a 5\% difference at $x=0.7$. We will be able to measure $a_1$ across this $x$ range with a statistical precision that is typically below 1\% and systematic uncertainties of 0.6-0.7\%.

\section{\label{sec:exp}Experimental Design and Projections}
\begin{figure}[htb]
    \begin{center}
        \includegraphics[width=0.66\textwidth]{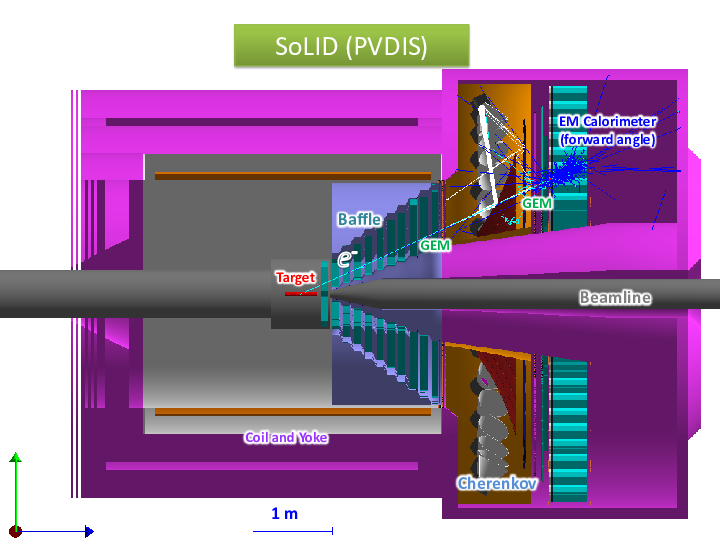}
        \caption{SoLID side-view of the Geant4 configuration with full detector setup.}
        \label{fig:des:solid_geant4}
    \end{center}
\end{figure}

The SoLID project is a large acceptance, high luminosity spectrometer and detector system designed for experiments that require a broad kinematic acceptance at high rates.  It presently has more than five approved experiments not counting parasitic running covering physics topics such as PVDIS on LD$_2$ and LH$_2$, semi-inclusive DIS on polarized targets, and $J/\psi$ production at threshold. This represents a major and unique Jefferson Lab physics program~\cite{JeffersonLabSoLID:2022iod,Achenbach:2023pba}, with a broadened physics program enabled by possible future JLab upgrades, e.g. positron beams or a future energy upgrade~\cite{Accardi:2023chb}.  

PVEMC will utilize the PVDIS configuration. This configuration with baffles nominally has symmetric azimuthal acceptance, polar angle acceptance of 22-35$^\circ$, and momentum acceptance of 1-7~GeV.  Azimuthally it is divided into 30~predominantly independent sectors which can operate at a total rate of $\sim$600~kHz in inclusive running.  A representation of this setup from our Geant4 simulation is shown in Fig.~\ref{fig:des:solid_geant4}. We anticipate about a total 155~kHz coincidence trigger using both the calorimeter and gas Cherenkov signal compared to the 500~kHz trigger for the PVDIS LD$_2$ measurement.  We will run the 30 sectors independently requiring about 5~kHz/sector for the primary measurement~\cite{JeffersonLabSoLID:2022iod}. 

\begin{figure}[htb]
    \begin{center}
        \includegraphics[width=0.7\textwidth]{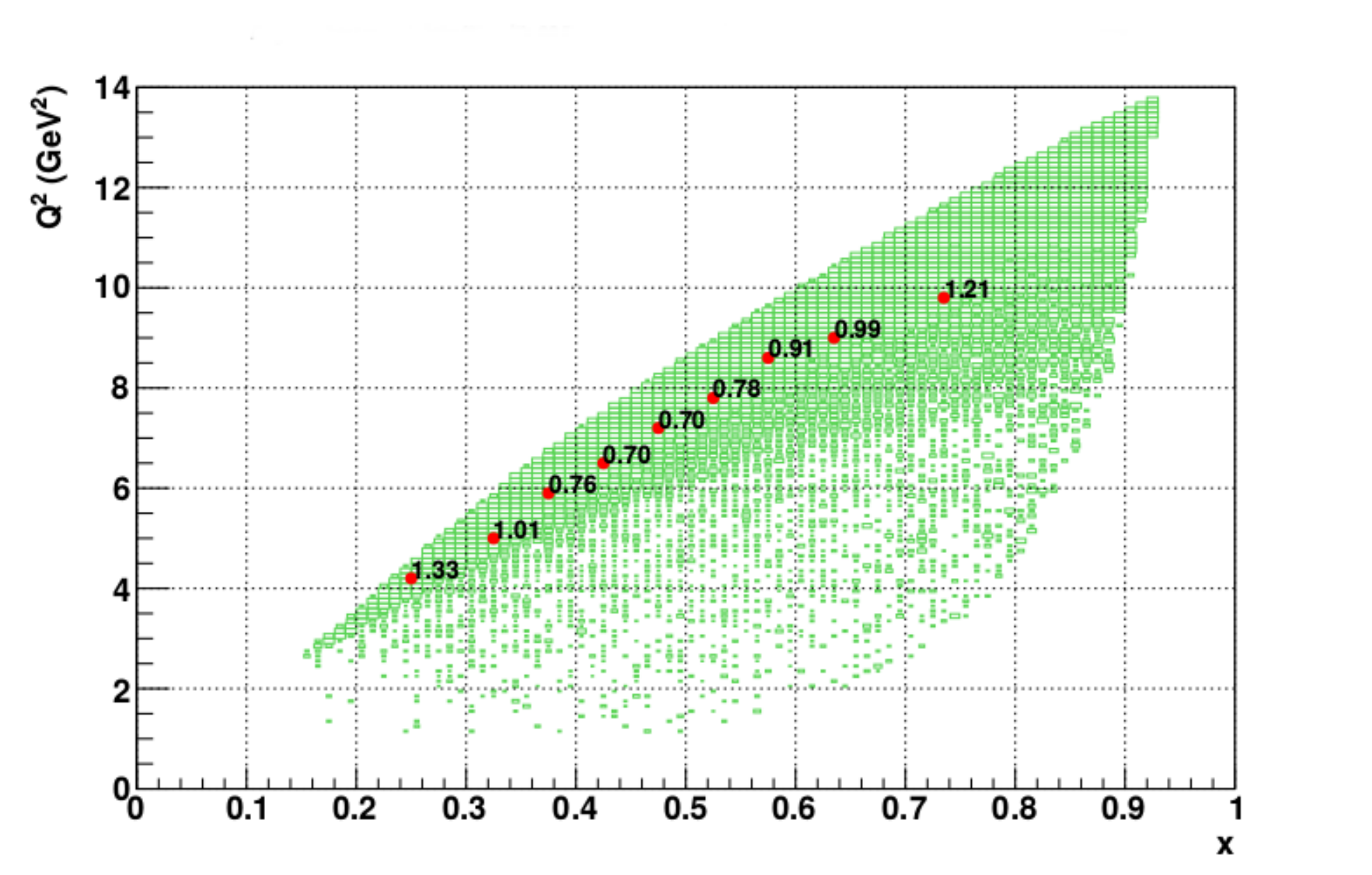}
        \caption{PVEMC statistical precision for $A_{PV}$ for $x$ and $Q^2$ bins in \%.}
        \label{fig:proj:Q2vsx}
    \end{center}
\end{figure}

\begin{figure}[htb]
    \begin{center}
    \includegraphics[height=0.48\textwidth, angle=90, trim=20mm 25mm 20mm 30mm, clip]{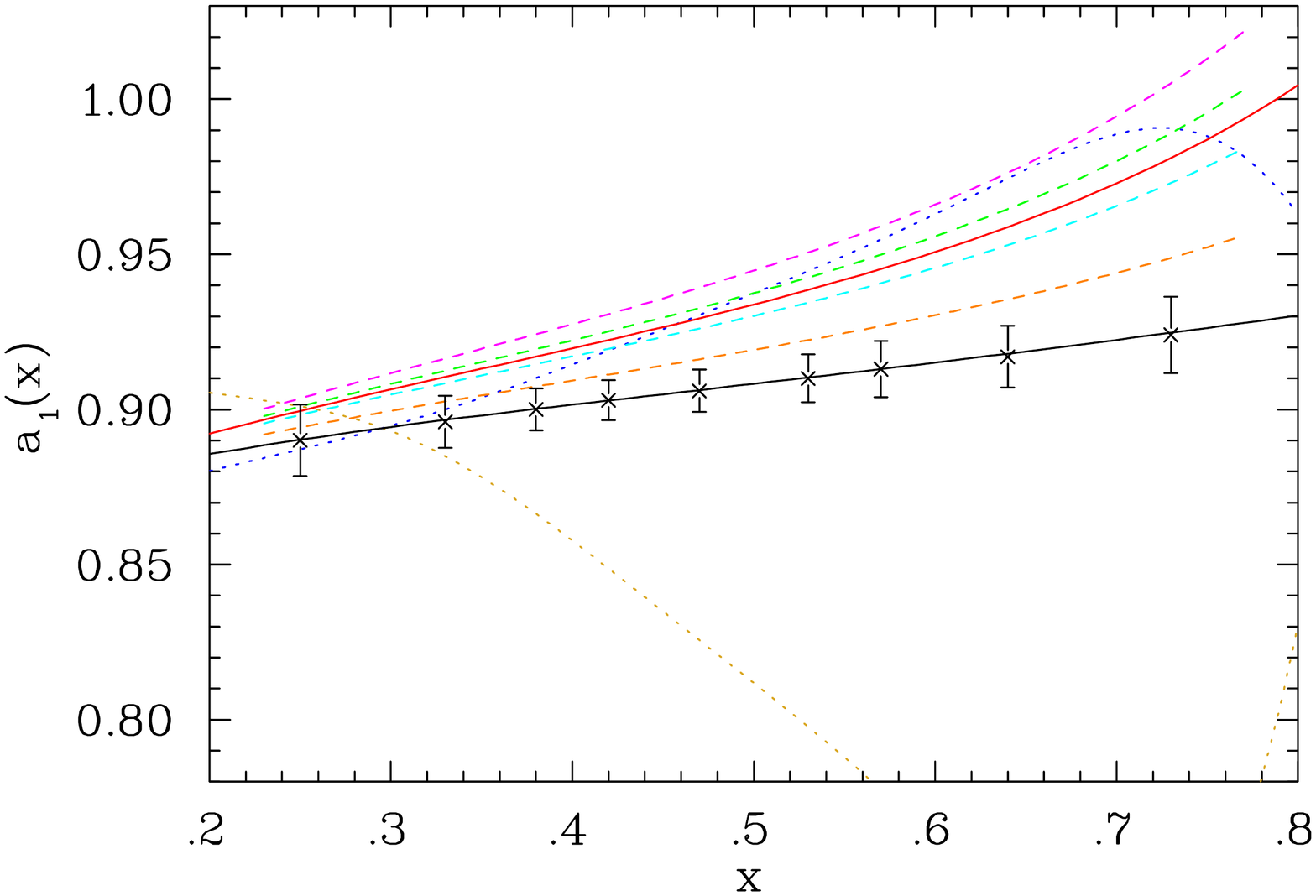}
    \includegraphics[height=0.48\textwidth, angle=90, trim=20mm 25mm 20mm 30mm, clip]{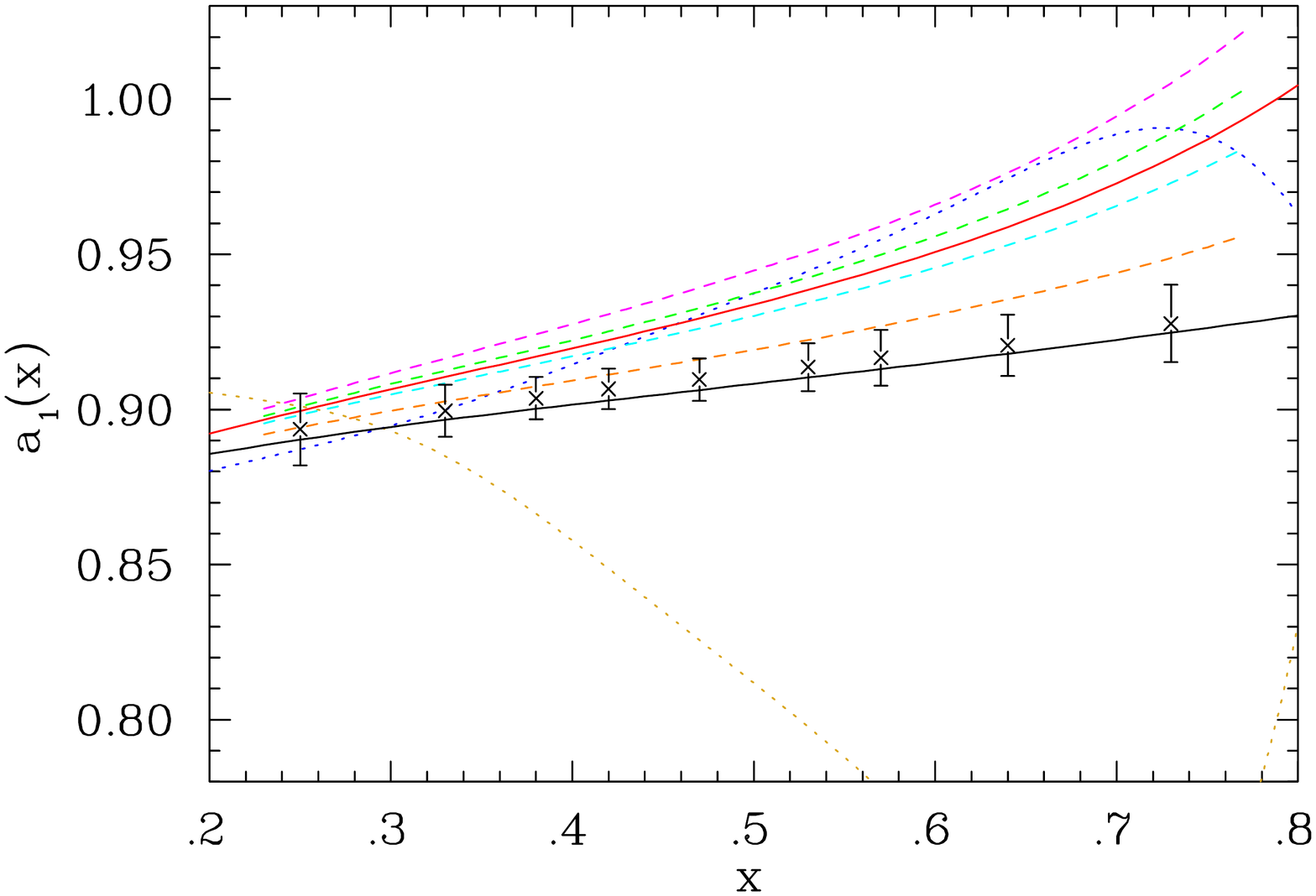}
        \caption{(Left) Projected $a_1$ data uncertainties (black points) generated assuming a flavor-independent EMC effect (solid black curve). The red line is the CBT model, the short-dashed lines assume the EMC effect occurs entirely within the up (blue) or down (brown) quarks, and the long-dashed curves correspond to the scaling predictions discussed in the PAC proposal~\cite{PAC50_PVEMC_Proposal}: from top to bottom the curves assume that the EMC effect scales with the fraction of high-momentum nucleons, the probability for two nucleons to be within 1~fm of each other, the average nucleon kinetic energy, and the average density seen by the protons and neutrons, based on calculations from Ref.~\cite{Arrington:2015wja}. (Right) Same, with the data shifted towards the CBT calculation by the 0.4\% to illustrate the impact of the normalization uncertainty}
        \label{fig:proj:a1}
    \end{center}
\end{figure}

We choose a $2.4~\mathrm{g/cm^2}$ $^{48}$Ca target, corresponding to 12\% of a radiation length, and an $80~\mathrm{\mu A}$ 11 GeV electron beam with maximum longitudinal beam polarization as the general running conditions.  The experimental layout we propose is identical to the existing SoLID PVDIS proposal as discussed in Ref.~\cite{JeffersonLabSoLID:2022iod}, with a $^{48}$Ca target system. For 68 days of data taking (assuming 100\% efficiency), a 0.7-1.3\% statistical uncertainty for $0.2<x<0.7$ can be obtained, as shown in Fig.~\ref{fig:proj:Q2vsx}.  A 0.6-0.7\% systematic uncertainty per-bin is anticipated, as discussed in Ref.~\cite{PAC50_PVEMC_Proposal}.  Our sensitivity to the CBT and a few other models of the flavor dependence are shown in Fig.~\ref{fig:proj:a1} and provides a $\sim$8$\sigma$ sensitivity for CBT model neglecting normalization uncertainty, or a 7$\sigma$ sensitivity if we arbitrarily apply a 0.4\% shift in the overall normalization, corresponding to the projected normalization uncertainty associated with beam polarization, to minimize the signal.

\begin{table}[htb]
    \begin{center}
        \begin{tabular}{lr}
            Effect & Uncertainty [\%] \\
            \hline
            Pions (bin-to-bin)  & 0.1-0.5 \\
            Charge-symmetric background & $<$0.1 \\
            Radiative Corrections (bin-to-bin) & 0.5-0.1 \\
            $R^{\gamma Z}/R^\gamma$ & 0.2 \\
            Other corrections including CSV    & 0.2 \\
            PDF uncertainties & 0.2 \\
            \hline
            \textbf{Total systematic \& model dependence} & \textbf{0.6-0.7} \\
            \hline
            \textbf{Statistics} & \textbf{0.7-1.3} \\
        \end{tabular}
    \caption{Summary of the systematic error contributions to our measurement after correcting for the pion and charge-symmetric backgrounds. The first three items are experimental uncertainties, while the remainder do not impact the uncertainty in $a_1$, but are relevant in the interpretation of the data in terms of flavor dependence of the EMC effect. We have included these uncertainties along with the experimental systematics for purposes of evaluating the sensitivity of the measurement. Note that an overall 0.4\% normalization, associated with the measurement of the beam polarization, is not listed in the table.}
    \label{tab:proj:sys}
    \end{center}
\end{table}

The total systematic uncertainties are summarized in Table~\ref{tab:proj:sys} and discussed further below.
Two independent polarimeters will be deployed for this experiment.  Intermittent measurements of the beam polarization will be made using a M\o ller polarimeter. These measurements are destructive to the beam so can only be made during dedicated measurements and must be made at low currents. Work is ongoing to improve the systematic uncertainty of the Hall A  M\o ller polarimeter to 0.4\%. Continuous monitoring of the polarization will be done by the upgraded Compton polarimeter, which is anticipated to give 0.4\% systematic uncertainty using both the photon and electron detectors. Details of the experimental systematic uncertainties are presented below, while estimates of the model dependence uncertainties are provided in the proposal~\cite{PAC50_PVEMC_Proposal}.

Our anticipated pion contamination to the electron signal is expected to be no worse than 4\% in any given bin based on the combined rejection factors in the Cherenkov and the preshower and shower counters. This is a small effect but still large enough to require a correction to the measured asymmetry and this requires good characterization of the pion contamination and the pion asymmetry. A fraction of beam time will be dedicated to characterizing this effect.

The charge symmetric background is about 15\% at x=0.25, but decreases rapidly as $x$ increases.  This background is not expected to carry a significant asymmetry, but the dilution must be precisely determined to minimize the uncertainty on the final asymmetry.  Dedicated measurements will be made with the solenoid magnet polarity reversed to measure the charge symmetric background. A detailed discussion of this background is available in the PAC proposal~\cite{PAC50_PVEMC_Proposal}. 

Based on the detailed projections in our PAC proposal~\cite{PAC50_PVEMC_Proposal}, we estimate the size of the radiative corrections to be $\ltorder$5\%, with the largest corrections occurring at the smallest $x$ values. We estimate that these corrections will be understood with $\approx$10\% relative uncertainty, and so we assign a 0.5\%-0.1\% bin-to-bin systematic error. Combining all of the experimental uncertainties, along with the estimated model-dependent contributions, we obtain a 0.6-0.7\% total systematic uncertainty across the entire $x$ range.

\section{Conclusions}\label{sec:conclusions}

We have presented a summary of the proposed parity-violating DIS measurement using the SoLID spectrometer and a $^{48}$Ca target. This provides a clean and a precise measurement of the flavor dependence of the EMC effect over the full $x$ range of the conventional EMC effect. This measurement will provide a completely new observable that can be used to differentiate between different underlying pictures of the physics driving quark modification in nuclei, yielding $\sim$8$\sigma$ sensitivity to the effects predicted in the CBT model~\cite{Cloet:2009qs}. In addition, it will provide constraints on the d- and u-quark distributions for heavy neutron-rich nuclei, which could impact the NuTeV anomaly, and provide a better understanding of the flavor structure of the pdfs in heavy nuclei, which are important input for high-$x$ measurements in A-A and e-A collisions and in $\nu$-A scattering measurements.

\begin{acknowledgements}
This work was supported by U.S. Department of Energy, Office of Science, Office of Nuclear Physics, under contract numbers DE-AC02-05CH11231 and DE-AC05-06OR23177.

\end{acknowledgements}


\bibliographystyle{spphys}       % APS-like style for physics
\bibliography{PVEMC_EPJA_arXiv}   % name your BibTeX data base

\end{document}